\documentclass[3p]{elsarticle}

\usepackage{lineno,hyperref}
\modulolinenumbers[5]

\usepackage{amsmath}
\usepackage[T1]{fontenc}

\journal{arXiv}

\begin{document}

\begin{frontmatter}

\title{Ion acceleration in non-relativistic quasi-parallel shocks using fully kinetic simulations}

\author[mps,nwu]{Cedric Schreiner}
\cortext[Cedric Schreiner]{Corresponding author}
\ead{mail@cschreiner.de}

\author[lanl,nwu]{Patrick Kilian}
\author[ita,nwu]{Felix Spanier}
\author[tub,mps]{Patricio A. Mu\~noz}
\author[tub,mps]{J\"org B\"uchner}

\address[mps]{Max-Planck-Institute for Solar System Research, Justus-von-Liebig-Weg 3, 37077 G\"ottingen, Germany}
\address[nwu]{Centre for Space Research, North-West University, 2520 Potchefstroom, South Africa}
\address[lanl]{T-2 Astrophysics and Cosmology, Mail Stop B283, P.O. Box 1663, Los Alamos National Laboratory, Los Alamos, NM 87545, USA}
\address[ita]{Zentrum f\"ur Astronomie der Universit\"at Heidelberg,
Institut f\"ur Theoretische Astrophysik, Albert-\"Uberle-Str. 2, 69120 Heidelberg, Germany }
\address[tub]{Technische Universit\"at Berlin, Zentrum f\"ur Astronomie und Astrophysik, Hardenbergstra{\ss}e 36, 10623 Berlin, Germany}

\begin{abstract}
The formation of collisionless shock fronts is an ubiquitous phenomenon in space plasma environments.
In the solar wind shocks might accompany coronal mass ejections, while even more violent events, such as supernovae, produce shock fronts traveling at relativistic speeds.
While the basic concepts of shock formation and particle acceleration in their vicinity are known, many details on a micro-physical scope are still under discussion.
In recent years the hybrid kinetic simulation approach has allowed to study the dynamics and acceleration of protons and heavier ions in great detail.
However, Particle-in-Cell codes allow to study the process including also electron dynamics and the radiation pressure.
Additionally a further numerical method allows for crosschecking results.
We therefore investigate shock formation and particle acceleration with a fully kinetic particle-in-cell code.
Besides electrons and protons we also include helium and carbon ions in our simulations of a quasi-parallel shock.
We are able to reproduce characteristic features of the energy spectra of the particles, such as the temperature ratios of the different ion species in the downstream which scale with the ratio of particle mass to charge.
We also find that approximately 12-15\% of the energy of the unperturbed upstream is transferred to the accelerated particles escaping the shock.
\end{abstract}


\end{frontmatter}


\section{Introduction}
\label{sec:introduction}

Collisionless shock fronts have been identified as one of the most prominent sites of astrophysical particle acceleration.
Mainly by the interaction with the magnetic fields in the transition region between upstream and downstream particles can be accelerated to relativistic energies (as first suggested by \cite{fermi_1949}), if they manage to cross the shock front several times before they escape.
The key mechanism in this scenario is diffusive shock acceleration (DSA) \cite{krymskii_1977,bell_1978,blandford_1978}, which assumes that upstream particles arriving at the shock may be reflected back into the upstream, where they interact with turbulent magnetic field fluctuations close to the shock.
As a result, they might be scattered back towards the shock, allowing the particles to cross the shock once more and to repeat the process.
Each shock crossing leads to a gain in energy, until some particles finally arrive at energies high enough to escape into the upstream.
For the case of (quasi-) perpendicular shocks, i.e.~shocks which propagate perpendicular to the ambient magnetic field in the upstream, an additional acceleration mechanism was suggested \cite{jokipii_1987}:
In the so-called shock drift acceleration (SDA), particles can gain additional energy by drifting along the shock front due to the electric field $-\vec{E} \propto \vec{v}_\mathrm{s} \times \vec{B}$ resulting from the motion of the shock with velocity $\vec{v}_s$ and the magnetic field $\vec{B}$.

Since the concept of DSA was first introduced, the theory of shock acceleration has evolved steadily (see e.g.~reviews by \cite{drury_1983,jones_1991,reames_1999,giacalone_2006,treumann_2009,desai_2016}) and is by now well established.
It has become clear that shock acceleration is a universal process that can be found anywhere from planetary bow shocks or coronal mass ejection (CME) driven shocks in the solar wind to massive, relativistic shocks expanding around supernova remnants (SNR).
However, there are some questions still under discussion, such as the injection problem.
It is generally assumed that a supra-thermal seed population is required for DSA to work efficiently, since only particles above a certain threshold energy can be injected into the acceleration process.
Recent models \cite{caprioli_2015} and spacecraft observations near the Earth's bow shock, supported by numerical simulations \cite{sundberg_2016}, suggest that a substantial fraction of thermal protons are reflected by the shock front, leading to an initial gain of energy.
Wave-particle scattering in the turbulent shock foot then allows some particles to return to the shock and to be further accelerated.
It was shown that only a few iterations are enough to sufficiently accelerate some particles to be able to enter the DSA process.

Apart from purely theoretical considerations and observations, numerical simulations have become an essential part in the investigation of shock acceleration.
Various simulation techniques are applied to study different details of shock evolution and particle acceleration.
In-situ \cite{riley_2016} or remote observations \cite{bemporad_2010,bemporad_2014} allow to reconstruct the plasma parameters near CME driven shocks.
The results can be the basis for subsequent magnetohydrodynamic (MHD) simulations of the macroscopic evolution of the shock \cite{riley_2016,bacchini_2015}.
On the other hand, test particle \cite{giacalone_2005,guo_2015} or Monte Carlo simulations \cite{vainio_2007,ng_2008} can be used to gain more insight into the micro-physical behavior of individual particles.

With the increase of computational resources available to researchers over the past years self-consistent, kinetic simulations have become more and more popular.
Currently, there are mainly two models for kinetic plasma simulations in use:
The fully kinetic model treats both electrons and ions as kinetic particles and allows to accurately model almost all micro-physical phenomena in a collisionless plasma.
The particle-in-cell method realizes this plasma model.
The hybrid model on the other hand only treats the ions as particles, whereas the electrons are modeled by a fluid.
Simulations employing the hybrid model use the PiC technique for ions and fluid techniques for the electrons.
Most hybrid simulations employ the Darwin approximation, which makes them essentially radiation-free.

The PiC method is mainly used to model shocks in pair plasmas \cite{bret_2017} or to focus on the dynamics of electrons, their injection into DSA, and the resulting energy spectrum \cite{matsukiyo_2006,matsukiyo_2010,matsukiyo_2012,matsukiyo_2015,guo_2017}.
Recently, the acceleration of both electrons and protons by means of DSA at a quasi-parallel shock has been simulated in a single, one-dimensional PiC simulation \cite{park_2015}.
Using a newly developed implicit PiC code, \cite{romanky_2017} have investigated the influence of alpha particles on the acceleration of electrons and protons in a one-dimensional setup.

However, as soon as the ion dynamics are in the focus of research, hybrid simulations are much more economic than fully kinetic PiC simulations.
The hybrid approach models the electrons by a fluid description, thus eliminating the need to resolve the comparably small time and length scales of the electrons, which are required for PiC.
Consequently, the computational cost per physical volume and time period is drastically reduced in a hybrid simulation compared to a corresponding run employing a PiC code.
Extended studies \cite{gargate_2012,gargate_2014} have shown that the acceleration efficiency for protons is higher at smaller shock angles $\theta_\mathrm{Bn}$, i.e.~the angle between the ambient magnetic field and the shock normal.
Comparisons of simulations carried out using one, two, or three spatial dimensions could demonstrate that two-dimensional simulations are sufficient to produce the correct micro-physical field structure at the shock front, while one-dimensional simulations lack some of the dynamic features and ripples at the shock, but generally still yield reasonable results \cite{guo_2013,caprioli_2014,hao_2017}.
However, only three-dimensional simulations are able to entirely reveal the complex interplay of small-scale fluctuations, waves and instabilities at the shock front\cite{burgess_2016}.

In addition to protons, heavier ions can also play a role in shock formation and particle acceleration.
For example, \cite{kropotina_2015} have performed three-dimensional hybrid simulations of SNR shocks driven solely by heavier ions.
With respect to the solar wind, the abundance of helium ions is known to noticeably influence the structure of CME shocks compared to shocks in a pure electron-proton plasma.
According to \cite{gedalin_2017} the presence of helium leads to a higher magnetic field strength and lower pressure in the downstream.
Preferential heating and acceleration of different ions at shock fronts is another topic still under discussion.
It is clear that the ratio of ion mass $A$ to charge $Q$ plays an important role for the efficiency of ion acceleration.
However, while measurements suggest that ions with larger $A/Q$ are accelerated less efficiently \cite{allegrini_2008}, recent two-dimensional hybrid simulations have shown the opposite effect \cite{caprioli_2017}.

The brief overview presented above makes clear that the acceleration of charged particles at collisionless shock fronts is an active field of research and that numerical simulations make crucial contributions to this topic.
They can help to clarify some of the longstanding problems, such as the injection problem, but also raise new issues, such as in the discussion of the $A/Q$ dependence of acceleration efficiency.
To assure quality results, it is important to rely not only on one numerical approach, but to reproduce results using a variety of methods.

In this article we investigate the acceleration of protons and heavier ions at a non-relativistic, quasi-parallel shock.
Although this topic is typically investigated using hybrid simulations, we choose the PiC method for our simulations.
We present the spectra of accelerated heavy ions (helium, carbon) resulting from a fully kinetic PiC simulation.
Furthermore we address the question of $A/Q$ dependence of acceleration efficiency by considering both He$^+$ and He$^{2+}$ ions.

The article consists of a short description of the numerical method and the setup used for the simulations in Sect.~\ref{sec:method}.
Results are then presented in Sect.~\ref{sec:results}.
Finally, a short discussion concludes the article in Sect.~\ref{sec:conclusions}.

\section{Simulation Method and Setup}
\label{sec:method}

We use the fully relativistic, explicit particle-in-cell (PiC) code ACRONYM \cite{kilian_2011}.
The code is available in 1d3v, 2d3v, and 3d3v versions, meaning that either one, two, or three spatial dimensions are resolved, while particle velocities, field, and current densities are always treated as three-dimensional vectors.
ACRONYM allows for various particle species, including electrons, protons, and heavier ions with arbitrary mass and charge.

In order to reduce the numerical effort, the shock simulations presented here are performed using the 1d3v version of the code.
The one-dimensional simulation box is oriented such that the $x$-direction is resolved.
A reflecting boundary is installed at $x=0$, while the opposite end of the simulation box is open, allowing particles to enter or escape.
In the $y$- and $z$-directions the simulation is invariant.

The simulation is initialized with a streaming plasma consisting of thermal electrons, protons, and a small fraction of heavier ions as detailed below.
The initial plasma stream represents the upstream and moves in $-x$-direction towards the reflecting boundary, where particles are reflected and heated while interacting with the incoming downstream plasma.
This produces the shock front, which then moves further into the upstream.
Thus, the simulation is carried out in the rest frame of the downstream plasma.

The shock angle $\theta_\mathrm{Bn}$ is chosen by adjusting the initial ambient magnetic field in the upstream, since the shock normal is fixed along the $x$-axis.
Considering \cite{gargate_2012}, who find that acceleration efficiency is best for $15^\circ < \theta_\mathrm{Bn} < 30^\circ$, we choose $\theta_\mathrm{Bn} = 20^\circ$ for our simulations.
We plan the setups for our simulations with CME driven shocks in the solar wind in mind.
However, the physical parameters used are not exactly representing the situation in the heliosphere, but are a compromise towards numerical efficiency and computational feasibility.
The general setup for all of our simulations employs the following physical parameters for the upstream plasma:
We set the ratio of the electron plasma frequency to the electron cyclotron frequency to $\omega_\mathrm{p,e} / |\Omega_\mathrm{e}| = 12.5$ and the thermal speed of the electrons to $v_\mathrm{th,e} = \sqrt{k_\mathrm{B} \, T / m_\mathrm{e}} = 0.04 \, c$, where $k_\mathrm{B}$ is the Boltzmann constant, $T$ is the temperature, $m_\mathrm{e}$ is the electron mass, and $c$ is the speed of light.
The resulting electron plasma beta is $\beta_\mathrm{e} = 0.50$.
The length of the simulation box is $L_x = 9.42 \cdot 10^3 \, \lambda_\mathrm{e}$, where $\lambda_\mathrm{e} = c / \omega_\mathrm{p,e}$ is the electron inertial length.
Note that we do not give the parameters for the protons here, since the proton density is different in different simulations, as will be listed below.
However, both protons and heavier ions have the same temperature as the electrons and the plasma is quasi-neutral.
The flow speed of the upstream plasma relative to the downstream is $v_\mathrm{u} = |\vec{v}_\mathrm{u}| = |(-8.27 \cdot 10^{-2} \, c, \, 0, \, 0)|$.
From solving the Rankine-Hugoniot jump conditions (e.g.~\cite{balogh_2013} chapter~3.3) for an electron-proton plasma (i.e.~without considering any heavier ions) we expect an Alfv\'enic Mach number of $M_\mathrm{A} = v_\mathrm{shock}/v_\mathrm{A} = 11.1$ (with the Alfv\'en speed $v_\mathrm{A}$) and a magnetosonic Mach number of $M_\mathrm{ms} = v_\mathrm{shock} / v_\mathrm{ms} = 9.3$ (with the speed of the magnetosonic mode $v_\mathrm{ms} = \sqrt{v_\mathrm{s}^2+v_\mathrm{A}^2}$, where $v_\mathrm{s}$ is the speed of sound).
This choice of the Mach number is mainly a result of numerical considerations, since a higher Mach number reduces the computational cost of the simulation.
Compared to the average events in the solar wind, where the Mach number is typically around $M_\mathrm{A} \sim 2$ and might reach values of $M_\mathrm{A} = 6$ at max \cite{lugaz_2015, riley_2016}, the Mach number in our simulations appears to be rather large.
However, it is also not entirely unrealistic, since some events also exhibit even larger Mach numbers (for example, \cite{kilpua_2015} report the case of $M_\mathrm{A} = 21$).

The numerical setup is described by the following parameters:
We choose \mbox{$N_x = 10^6$} as the number of numerical grid cells in the simulation box and \mbox{$N_t = 10^7$} as the number of time steps.
The length of a time step is given by \mbox{$\Delta t = 5.44 \cdot 10^{-3} \omega_\mathrm{p,e}^{-1}$} and the grid spacing is \mbox{$\Delta x = 9.42 \cdot 10^{-3} \, c \, \omega_\mathrm{p,e}^{-1}$}, which means that the Debye length in the upstream is resolved by about three grid cells.
We artificially reduce the proton mass to $m_\mathrm{p} = 64 \, m_\mathrm{e}$, where $m_\mathrm{e}$ is the natural electron mass.
The masses $m_\mathrm{i}$ of heavier ions are scaled down accordingly, maintaining the natural $m_\mathrm{i} / m_\mathrm{p}$ ratios.
The simulations employ 64 particles per species per cell, i.e.~about 200 particles per species per Debye length in the upstream and even more particles per Debye length in the downstream, because of the larger Debye length and the higher particle count per cell.
However, numerical particles of different species are initialized with different macro factors, meaning that they represent different numbers of physical particles.
This allows to adjust the physical number densities of protons and heavier ions while the number of numerical particles stays the same.
It also means that the plasma stays quasi-neutral, although there are twice as many numerical particles with positive charge than particles with negative charge.

We carried out three simulations, labeled S1, S2, and S3.
At initialization, each simulation features a population of electrons as described above, as well as a proton and a heavier ion population.
The additional heavier ions are helium (S1 and S2) and carbon (S3).
For their physical number densities relative to the protons we have chosen the approximate abundances measured in the solar wind (e.g.~\cite{gloeckler_1986,lodders_2003}).
Table \ref{tab:parameters} lists the properties of the proton and ion populations in each simulation.
Note that we are using the same number densities for He$^{2+}$ and He$^+$, although the abundances for these ions are of course different in the solar wind.

\begin{table}[htb]
	\begin{tabular}{c c c c c c c}
		\hline
		simulation & ion species & $A/Z$ & $n_\mathrm{i} / n_\mathrm{p}$ & $\lambda_\mathrm{i} / \lambda_\mathrm{e}$ & $\lambda_\mathrm{p} / \lambda_\mathrm{e}$ & $\beta$\\
		\hline
		S1 & He$^{2+}$ & $2$ & $5.0 \cdot 10^{-2}$ & $37.5$ & $8.4$ & $0.97$\\
		S2 & He$^{+}$ & $4$ & $5.0 \cdot 10^{-2}$ & $75.0$ & $8.2$ & $1.00$\\
		S3 & C$^{6+}$ & $2$ & $5.0 \cdot 10^{-4}$ & $206.6$ & $8.0$ & $1.00$\\
		\hline
	\end{tabular}
	\caption{
		Simulation parameters in the upstream plasma:
		ion species and charge, mass to charge ratio $A/Z$ of the ions in units of proton mass and charge, number density $n_\mathrm{i}$ of the ions relative to the protons, inertial lengths $\lambda_\mathrm{i}$ and $\lambda_\mathrm{p}$ of the ions and the protons relative to the electron inertial length $\lambda_\mathrm{e}$, and the plasma beta $\beta$ of the whole plasma (considering all three particle species).
		Note that the plasma beta results from the physical parameters chosen for the electrons (density, temperature) and the condition of charge neutrality (which then fixes the parameters for the ions).
		Setting the same parameters for the electrons in all simulations keeps the numerical quantities -- such as grid spacing and the length of a time step -- the same in all simulations as well.
		However, it leads to a plasma beta slightly below 1 in simulation S1.
	}
	\label{tab:parameters}
\end{table}

\section{Simulation Results}
\label{sec:results}

\subsection{Qualitative Phase Space Observation}
\label{sec:results_sub1}

\begin{figure}[p]
      \centering
      \includegraphics[width=0.9\linewidth]{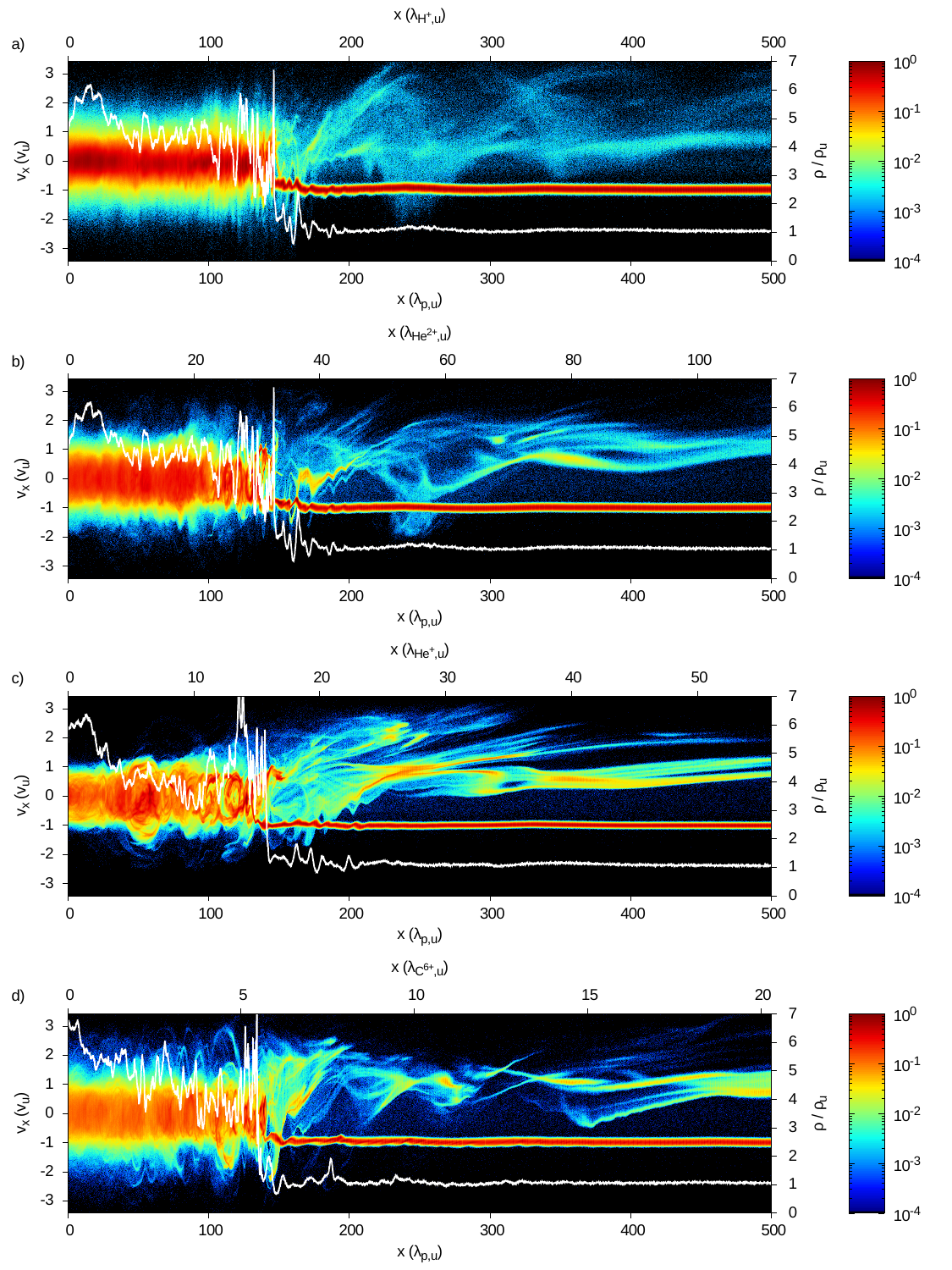}
      \caption{
      Phase space density of ions at $t \, \Omega_\mathrm{p} = 68.0$: a) H$^{+}$ ($A/Z=1$, i.e.~regular protons) from simulation S1, b) He$^{2+}$ ($A/Z=2$) from S1, c) He$^{+}$ ($A/Z=4$) from S2, and d) C$^{6+}$ ($A/Z=4$) from S3.
      Color coding shows the phase space density relative to the maximum in each panel.
      The white solid line represents the total plasma density $\rho(x)$ in units of the density $\rho_\mathrm{u}$ of the undisturbed upstream (see axis on the right-hand side).
      }
      \label{fig:phase_space}
\end{figure}

To illustrate the behavior of particles close to the shock front we show phase space diagrams from simulations S1, S2, and S3 in Fig.~\ref{fig:phase_space}.
The figure shows snapshots of the particle densities of protons (panel a), He$^{2+}$ (panel b), He$^{+}$ (panel c), and C$^{6+}$ (panel d) in $x$-$v_x$-phase space at $t \, \Omega_\mathrm{p} = 68.0$, i.e.~at the end of the simulation runs.
Looking at the downstream at the left side of each panel it can be seen that the particles isotropize close to the shock front for small $A/Z$, while ions with high mass to charge ratios develop structures that reach far into the downstream.
This behavior is both expected and consistent with hybrid simulations, such as those by \cite{caprioli_2017}.

On the upstream side of the shock a shock foot is established.
However, apart from fluctuations close to the shock the upstream plasma is propagating towards the shock (negative $v_x$) without further perturbations.
Additionally to the thermal bulk flow the upstream also contains particles propagating away from the shock, i.e.~with positive $v_x$.
These ions are reflected and accelerated by the shock and are thus able to outrun the shock front and escape into the upstream.
The reflected particles form filamentary structures in the upstream, with the precise appearance of the filaments depending on the $A/Z$ ratio:
For higher $A/Z$ the filaments are narrow and dense with comparably little variation in speed (see panel c).
For smaller $A/Z$, however, particle velocities are widespread (most notably seen for the protons in panel a) and the filaments vary in speed and density depending on the distance to the shock, following a wave-like shape.
Again, the behavior is consistent with results from hybrid simulations, as can be seen in Fig.~4 of \cite{caprioli_2017}.

The white lines in the individual panels of Fig.~\ref{fig:phase_space} show the plasma density $\rho(x)$ as a fraction of the density $\rho_\mathrm{u}$ of the undisturbed upstream.
While the density of the upstream (including reflected particles) is relatively constant, there are strong fluctuations in the downstream and around the shock itself.
The latter are expected and correspond to the shock foot, the initial overshoot at the shock front, and the downstream region close to the shock, where the plasma has not yet isotropized.
However, we also find strong fluctuations and high density ratios $\rho / \rho_\mathrm{u} \simeq 6$ far in the downstream, close to the reflecting wall which forms the end of the simulation box at $x = 0$.
These features are related to the numerical boundary conditions.
We thus exclude the plasma volume in the range $0 \leq x / \lambda_\mathrm{p,u} \leq 50$ (with the proton inertial length $\lambda_\mathrm{p,u}$ in the upstream) from our further analysis.

\subsection{Downstream Spectra}
\label{sec:results_sub2}

\begin{figure}[htb]
      \centering
      \includegraphics[width=0.8\linewidth]{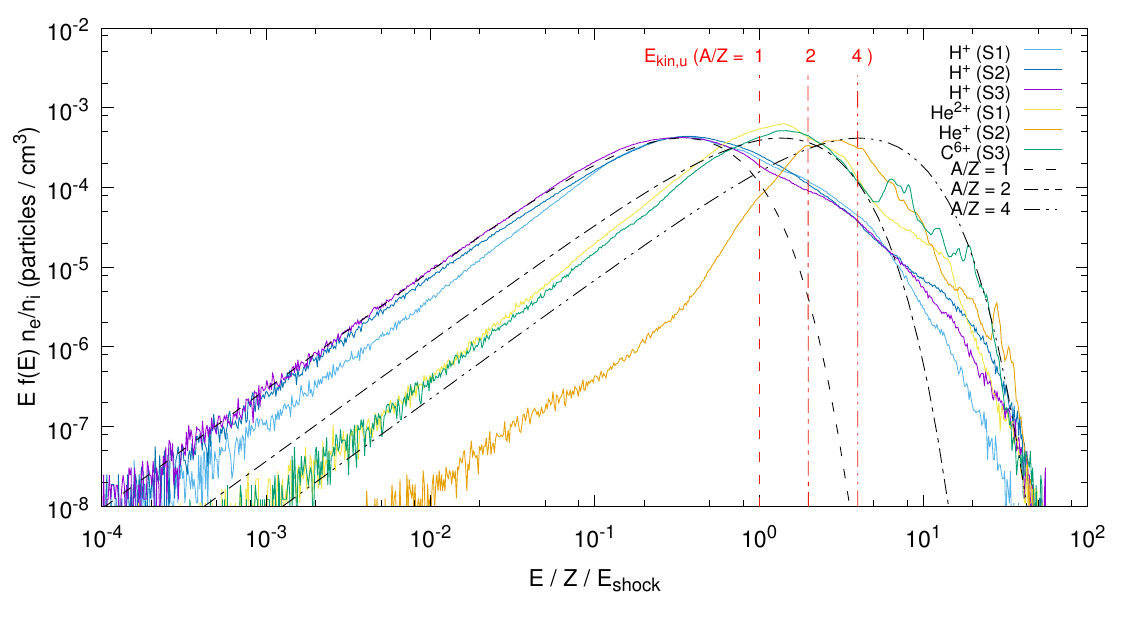}
      \caption{
      Downstream energy spectra of different ions from simulations S1, S2, and S3 at $t \, \Omega_\mathrm{p,u}= 68.0$.
      The H$^+$ ions are the regular protons in the simulation and are not meant to represent an additional ion species.
      Particle densities are normalized to the electron density $n_\mathrm{e}$ in the upstream, the energy is normalized to the charge $Z$ of the ions in units of the proton charge and the shock energy $E_\mathrm{shock} = 0.5 \, m_\mathrm{p} \, v_\mathrm{u}^2$.
      The dashed black line shows a fit to the thermal protons in the downstream.
      The dash-dotted and dash-dot-dotted black curves are derived from this fit by assuming a linear dependence of temperature and ion mass $A$ in units of the proton mass (similar to findings by e.g.~\cite{caprioli_2017}).
      Note that the maximum energy reached by the particles scales with the ion charge $Z$ in units of the proton charge, i.e.~that the cut-off occurs at the same $E/Z$ for all species.
      The vertical red lines mark the kinetic energy of the incoming ions with mass to charge ratios $A/Z = \{1,2,4\}$ as labeled.
      }
      \label{fig:downstream_spectra}
\end{figure}

We produce the energy spectra of the ions in the downstream at the end of our simulations at $t \, \Omega_\mathrm{p,u} = 68.0$.
The result is shown in Fig.~\ref{fig:downstream_spectra}, where the particle density is plotted over the particle energy $E$ normalized to the charge $Z$ and the shock energy $E_\mathrm{shock} = 0.5 \, m_\mathrm{p} \, v_\mathrm{u}^2$.
The behavior of the protons is very similar in all three simulations.
While the protons follow a thermal spectrum at lower energies, an excess of high energy particles can be seen at higher energies.
Clearly, supra-thermal protons are produced in the downstream, although the expected power law is not yet established.
The maximum energy reached by the protons in the downstream increases linearly with time throughout the simulation, as predicted for diffusive shock acceleration (e.g.~\cite{drury_1983,gargate_2012,caprioli_2014b}).
We thus expect that more energetic particles would be produced if the simulation went on and that a power law would finally be achieved.
However, with limited computational resources a continuation of the simulations over several hundred proton cyclotron time scales -- as is common in hybrid simulations (see e.g.~\cite{caprioli_2014} for an extensive study) -- is not feasible.

Comparing the different types of ions we find that the shape and extend of the energy spectra depends on the mass over charge ratio $A/Z$, as expected.
Thus, the spectra of He$^{2+}$ and C$^{6+}$ ions (both with $A/Z = 2$) are similar, while the He$^+$ spectrum ($A/Z=4$) differs.
It can also be seen in Fig.~\ref{fig:downstream_spectra} that the heavier ions have not yet fully relaxed into a thermal spectrum, which is most obvious for He$^+$.
Here the spectrum peaks at an energy slightly below the kinetic energy $E_\mathrm{kin,u}^{\mathrm{He}}$ of the incoming He$^+$ ions from the upstream.
These particles penetrate far into the downstream, as seen in Fig.~\ref{fig:phase_space}~d), and make up a large fraction of the He$^+$ population there.

Supra-thermal He$^{2+}$ and C$^{6+}$ ions can be found, although no power law trend can yet be seen.
Due to the high $A/Z$ ratio for He$^+$ hardly any non-thermal, energetic particles can be found for this species.
However, as discussed above for the protons, we expect that more energetic particles would be produced later on in the simulation.
It is also worth noting that the highest $E/Z$ reached by the particles is the same for all ion species.
Thus, the maximum energy scales with the ion charge $Z$.
Following the argumentation of \cite{caprioli_2014b}, who give the maximum energy reached by a particle based on an estimate of diffusion coefficients and acceleration time scales, this behavior is expected for diffusive shock acceleration.

Taking a closer look at the thermal parts of the spectra we are able to reproduce another typical feature.
We fit a thermal spectrum to the proton data (see dashed black line in Fig.~\ref{fig:downstream_spectra}).
We then scale the temperature up according to the mass to charge ratios of the heavier ions and plot the corresponding thermal spectra (dash-dotted and dash-dot-dotted black curves).
These match well with the data for the He and C ions, indicating an $A$ dependence of the ion temperature in the downstream.
Similar findings are reported from hybrid simulations (see e.g.~\cite{caprioli_2017}~Fig.~1).
The mass dependence of the temperature is qualitatively explained by the fact that all ions coming from the upstream arrive at the shock front with the same (bulk) velocity.
However, their kinetic energy depends on their mass, which -- assuming that the kinetic energy related to the bulk flow of the upstream is converted into heat -- is why heavier ions end up with a larger amount of thermal energy in the downstream.

We show the downstream energy spectra of the electrons in \ref{app:electrons}.

\subsection{Energy Composition and Upstream Spectra}
\label{sec:results_sub3}

We analyze the energy composition of the upstream and downstream plasma.
In general, we distinguish three types of energy:
We measure the energy density $w_\mathrm{em}$ of the electromagnetic fields, the energy density $w_\mathrm{part} (v_x\!<\!0)$ of the particles propagating in negative $x$-direction, and the energy density of particles traveling in the opposite direction ($v_x > 0$).

\begin{figure}[htb]
      \centering
      \includegraphics[width=0.8\linewidth]{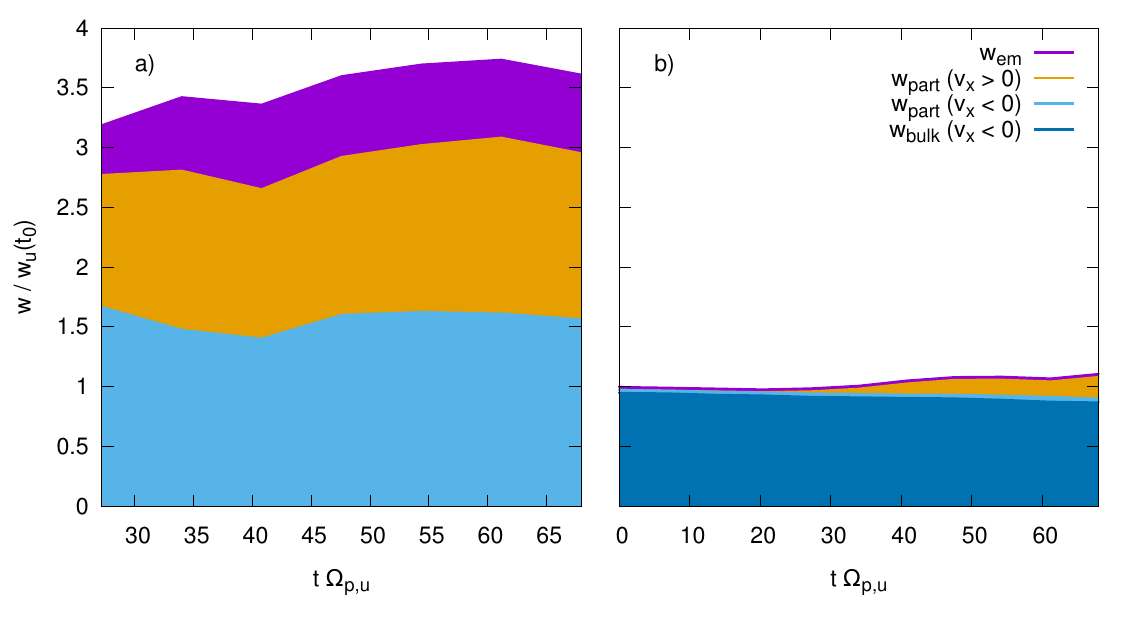}
      \caption{
      Energy composition of the downstream (panel a) and a region in the upstream (panel b) as functions of time.
      For the upstream the plasma volume between $159.0 \, \lambda_\mathrm{p,u}$ and $238.5 \, \lambda_\mathrm{p,u}$ from the shock front is considered.
      The energy density is normalized to the energy density $w_\mathrm{u} (t_0)$ of the upstream at the beginning of the simulation.
      Data is taken from simulation S1, but similar results can be obtained from S2 and S3 as well.
      We distinguish between field energy density $w_\mathrm{em}$ and the energy density $w_\mathrm{part}$ of the particles.
      Particles are further subdivided into those traveling in the direction of the upstream ($v_x\!<\!0$) and those propagating the other way ($v_x\!>\!0$).
      Since the simulation is performed in the rest frame of the downstream there is no bulk motion there.
      However, the upstream plasma is moving relative to the downstream and we can measure the energy density $w_\mathrm{bulk}$ related to this motion (see panel b).
      }
      \label{fig:energy_composition}
\end{figure}

In the reference frame of the simulation the downstream is at rest and the particles isotropize over time.
That means that the energy densities of particles propagating in positive and negative $x$-direction equalize, as can be seen in Fig.~\ref{fig:energy_composition}~a).
Towards the end of the simulation both the energy density of all particles with negative $v_x$ and the energy density of all particles with positive $v_x$ correspond to roughly 1.5 times the total energy density of the unperturbed upstream.
We also find a substantial fraction of electromagnetic energy in the downstream, which is related to the excitation of large scale plasma waves.
While there are also fluctuations in the electric field, the the energy related to magnetic field fluctuations clearly dominates.

In the upstream, however, the situation is different, as depicted in Fig.~\ref{fig:energy_composition}~b).
Since the upstream is moving relative to the stationary downstream, the largest contribution to the total energy density comes from the kinetic energy of the bulk motion.
We therefore distinguish between the energy density $w_\mathrm{bulk}$ related to the upstream flow and the remaining energy density $w_\mathrm{part} (v_x\!<\!0)$ of the particles propagating towards the shock, i.e.~mainly the thermal energy of the upstream particles (for this diagnostic all particles species are considered).
We measure these components by determining the total energy density $w_\mathrm{part}(v_x\!<\!0)$ of all particles with $v_x<0$ and the average velocity of the bulk flow (i.e.~mainly the protons, which contribute the most momentum).
From the average velocity we can derive an estimate for the bulk-kinetic energy density of the upstream.
Subtracting this quantity from $w_\mathrm{part}(v_x\!<\!0)$ gives us an estimate of the thermal energy of the downstream particles (with $v_x<0$).
The electromagnetic fields in the upstream are comparably weak, with the background magnetic field being the main contributor to $w_\mathrm{em}$.

While at the start of the simulation virtually all particles in the upstream propagate in negative $x$-direction, particles are reflected and accelerated once the shock has been established.
Thus, over time the reflected particles ($v_x > 0$) escape into the upstream (see Fig.~\ref{fig:energy_composition}~b).
By measuring the energy density $w_\mathrm{part}(v_x\!>\!0)$ of the reflected particles relative to the energy density of the unperturbed upstream, we find an acceleration efficiency of $12 - 15\%$.
We note that the largest contribution to the energy density $w_\mathrm{part}(v_x\!>\!0)$ comes from the protons.
While heavier ions are also reflected and accelerated (see Fig.~\ref{fig:phase_space}), they are less abundant and thus contribute less energy.
The contribution of electrons is also small, because most of the reflected electrons do not penetrate far into the upstream and the remaining few carry comparably little kinetic energy due to their low mass.

We further notice that the bulk flow in the upstream gradually slows down over time, as can be seen in Fig.~\ref{fig:energy_composition}~b).
The slowdown progresses steadily over time at a constant rate depending on the distance from the shock front.
To quantify the slowdown we subdivide the upstream into six sections of equal width $\Delta x = 79.5 \, \lambda_\mathrm{p,u}$, starting at the position of the shock front (which we will refer to as $x = 0$ in the following) and advancing further into the upstream (towards positive x).
In each output time step we determine the average speed $v_{x,\mathrm{bulk}}$ (measured in the downstream rest frame) of the mass flow in each section.
Since the main contribution to the mass flow comes from the protons, this quantity is mostly the same as the bulk flow speed of the upstream protons.

\begin{figure}[htb]
      \centering
      \includegraphics[width=0.8\linewidth]{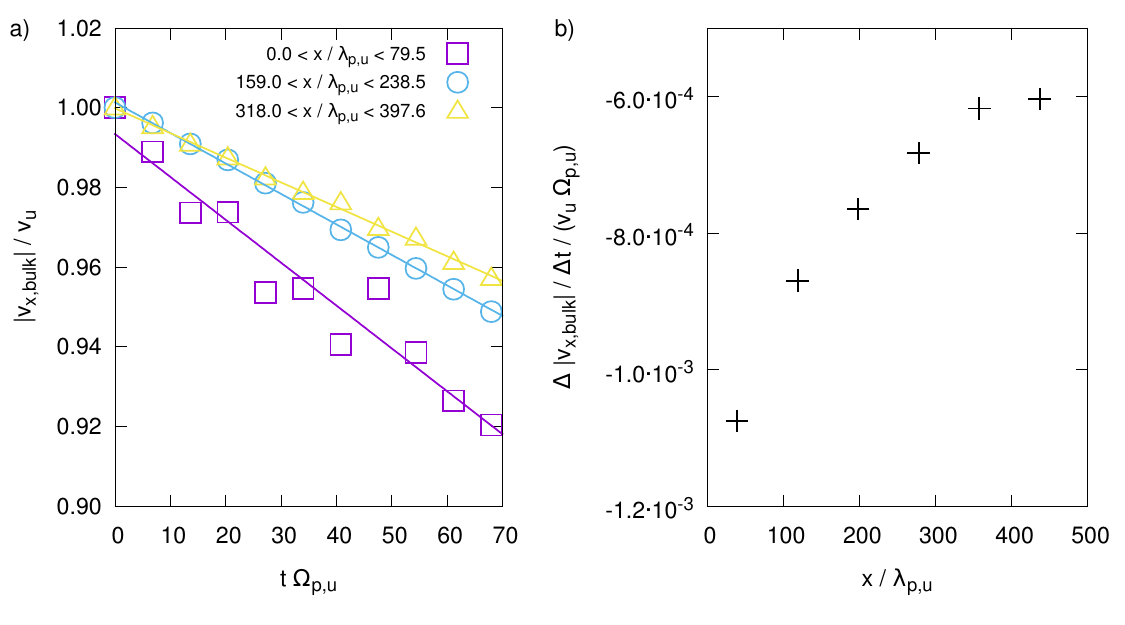}
      \caption{
      Panel a) shows the bulk flow speed $v_{x,\mathrm{bulk}}$ (measured in the downstream rest frame) of the upstream plasma in units of the initial speed $v_\mathrm{u}$ of the upstream.
      Data is shown for three sections of the upstream at different distances $x$ from the shock front as indicated in the key.
      Solid lines in the same colors represent linear fits to the data which are used to obtain the rate of the slowdown $\Delta v_{x,\mathrm{bulk}}/\Delta t$ of the upstream flow.
      Panel b) shows the rate of slowdown as a function of the distance from the shock, using data from a total of six sections in the upstream.
      The $x$-positions of the data points mark the middle of each section and are given relative to the position of the shock front.
      }
      \label{fig:slowdown}
\end{figure}

We show the time development of the bulk flow speed $v_{x,\mathrm{bulk}}$ in three of the six sections of the upstream in Fig.~\ref{fig:slowdown}~a).
In each section the bulk flow speed is a linear function of time.
We therefore apply a linear fit to our data and obtain the rate $\Delta v_{x,\mathrm{bulk}} / \Delta t$ describing the slowdown of the upstream flow.

With $\Delta v_{x,\mathrm{bulk}} / \Delta t$ being known in each section we can then plot the slowdown as a function of the distance $x$ to the shock front, as presented in Fig.~\ref{fig:slowdown}~b).
As expected, the rate at which the upstream slows down is largest in the section closest to the shock, presumably due to the interaction with the shock and particles from the downstream.
Further away from the shock the effect becomes weaker, i.e. the bulk plasma maintains more of its kinetic energy.
However, our data indicates that even at large distances from the shock the rate of the slowdown remains finite.
This hints at a mechanism that effectively takes away kinetic energy from the upstream even before it encounters the shock.
In Fig.~\ref{fig:energy_composition}~b) it can be seen that the bulk-kinetic energy decreases even before the first reflected particles enter the respective plasma volume at $t \, \Omega_\mathrm{p,u} \sim 25$.
At early times the energy is therefore entirely removed from the upstream (i.e. not just transferred between bulk-kinetic and thermal or field energy), while at later times a direct transfer of energy from particles with $v_x<0$ to reflected particles with $v_x>0$ in the same plasma volume might be possible.
We assume that even in the early stages the energy is transferred to the downstream or to reflected and accelerated particles moving away from the shock back into the upstream.
However, we could not find direct evidence of any such mechanism in our data.

To rule out numerical reasons for the slowdown we ran different test simulations:
First, we checked the influence of the number of particles per cell in test simulations with less time steps.
With all other physical and numerical parameters fixed we increased or decreased the number of numerical particles per cell.
In both cases the simulations reproduced the slowdown with no significant differences to the findings presented above.
In a second test we changed the boundary conditions of the simulation box.
Instead of the combination of one reflecting wall and one open end in $x$-direction we inserted periodic boundaries.
Thus, we could observe the drifting upstream plasma without the influence of the shock front.
In this test case the plasma maintained its drift speed.
A fundamental numerical error is thus unlikely, since an effect caused by e.g.~numerical collisions between particles, a loss of energy due to grid Cherenkov radiation, or numerical current filtering would persist in the setup with periodic boundary conditions.

Using the data shown in Fig.~\ref{fig:slowdown} we are able to perform a Galilean transformation into the rest frame of the upstream.
We use the bulk flow speed $v_{x,\mathrm{bulk}}(x,t)$ -- depending on both distance $x$ from the shock and time $t$ -- as the boost speed.
Thus, we have to perform individual Galilean transformations for each section of upstream plasma.
We can then analyze the particle data in each of the sections.

\begin{figure}[h!]
      \centering
      \includegraphics[width=0.5\linewidth]{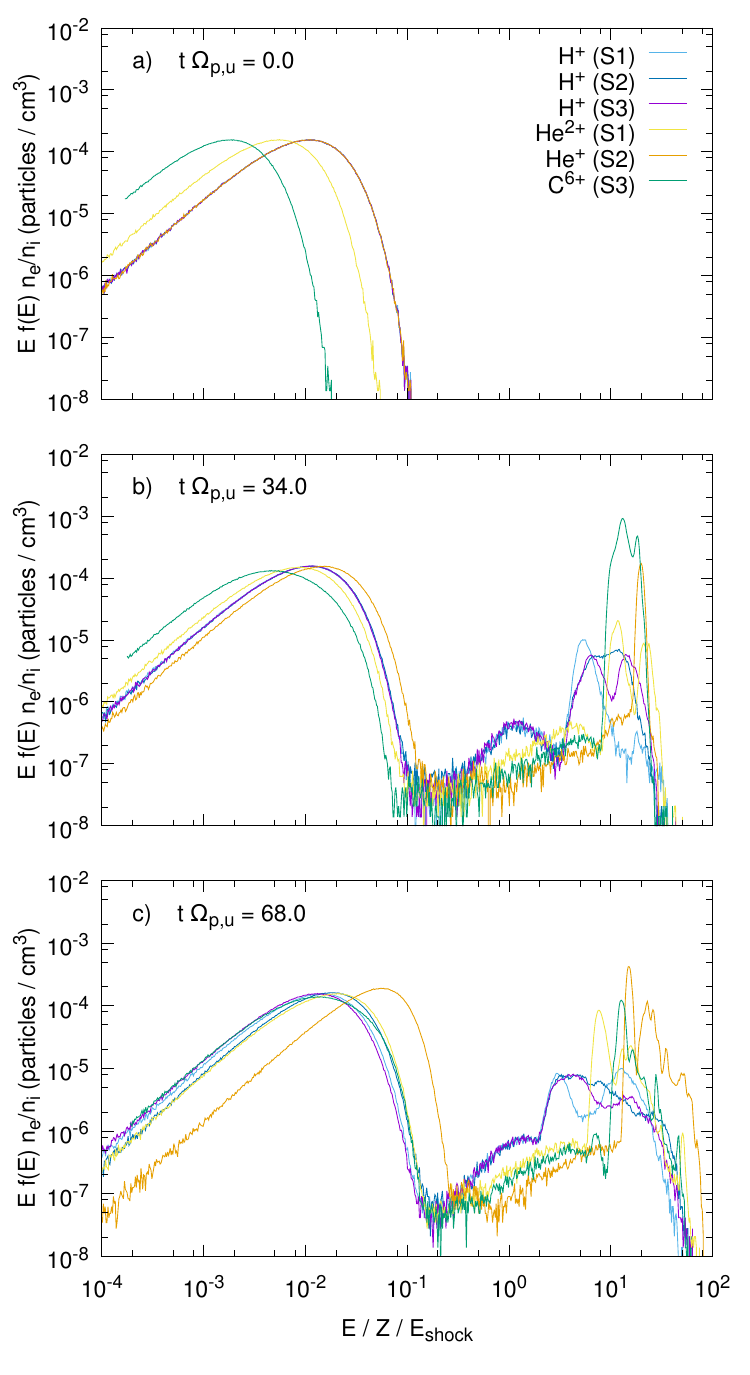}
      \caption{
      Upstream energy spectra of different ions from simulations S1, S2, and S3 at various points in time as labeled in the panels.
      Particle data was gathered in a spatial interval between $159.0 \, \lambda_\mathrm{p,u}$ and $238.5 \, \lambda_\mathrm{p,u}$ from the shock front.
      The energy is given in the rest frame of the unperturbed upstream.
      See text for a detailed description of the plots.
      }
      \label{fig:upstream_spectra_t}
\end{figure}

\begin{figure}[h!]
      \centering
      \includegraphics[width=0.5\linewidth]{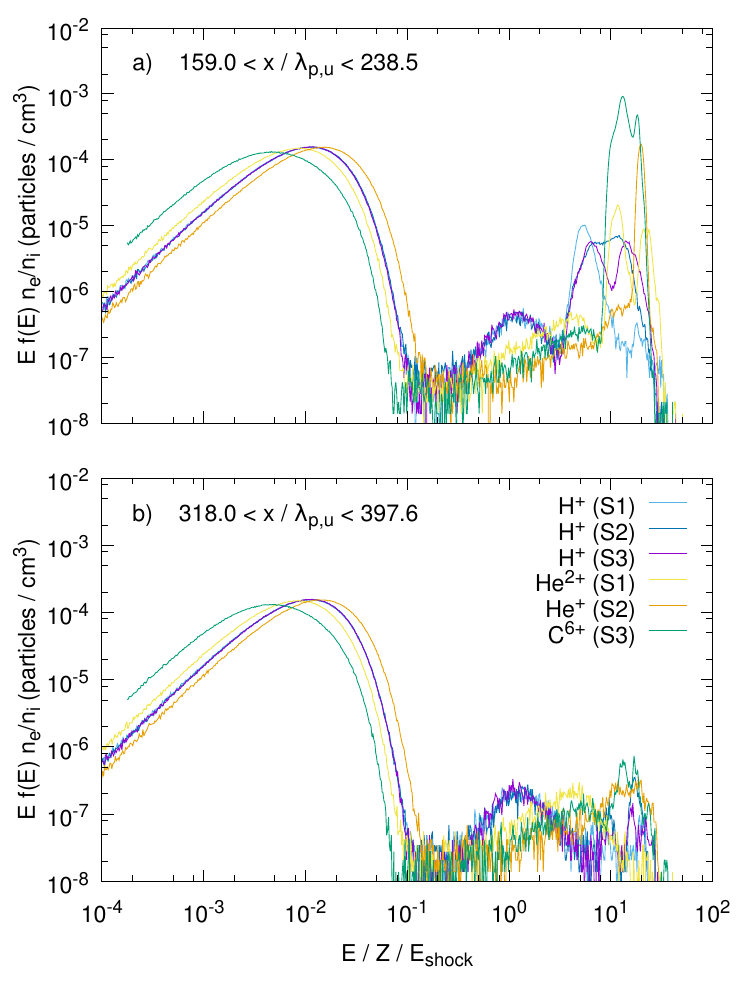}
      \caption{
      Upstream energy spectra of different ions from simulations S1, S2, and S3 at $t \, \Omega_\mathrm{p} = 34.0$ and two different distances from the shock front as labeled in the panels.
      See text for a detailed description of the plots.
      }
      \label{fig:upstream_spectra_x}
\end{figure}

We show energy spectra of the ions in the upstream in Figs.~\ref{fig:upstream_spectra_t} and~\ref{fig:upstream_spectra_x} and choose the rest frame of the bulk mass flow as our frame of reference.
Snapshots at a constant distance from the shock front but different points in time are depicted in Fig.~\ref{fig:upstream_spectra_t}.
Panel a) shows the thermal spectra of the upstream ion populations at the beginning of the simulation.
Note that all ion species have the same temperature, leading to the spectra of He$^{2+}$ and C$^{6+}$ being shifted to lower energies by a factor of their respective $Z$ compared to the protons.
At later times the thermal spectra of protons and heavier ions are shifted relative to each other, which is a result of the chosen rest frame.
While the bulk mass flow (predominantly made up by protons) slows down, the heavier ions maintain a larger fraction of their initial kinetic energy, thus propagating relative to the bulk flow.
This leads to He and C ions having a higher energy in the chosen rest frame of the bulk mass flow of the upstream.

In panels b) and c) of Fig.~\ref{fig:upstream_spectra_t} reflected ions can be found at energies larger than the typical thermal energy of the unperturbed upstream.
This second population of particles consists of a fraction of reflected, but mainly thermal particles (most noticeably seen for the protons around $E/E_\mathrm{shock} \sim 1$ in panel b) and a number of strongly accelerated particles at even higher energies.
We note an increase of the maximum energy reached by the reflected and accelerated ions.
This observation is in accordance with the downstream energy spectra (see Fig.~\ref{fig:downstream_spectra}), where the maximum particle energy is also increasing over time.

The high energy ions form a rather narrow distribution, as can be seen in Fig.~\ref{fig:upstream_spectra_t} or \cite{fiuza_2012} Fig.~1\,h).
This distribution is produced close to the shock front and requires some time to be built up.
These findings are in agreement with the idea that particles have to be accelerated by multiple shock crossings until they can escape from the shock region and propagate into the upstream.

In contrast to these high energy particles a population of reflected, but mainly thermal particles is created almost instantly upon contact with the shock.
Figure~\ref{fig:upstream_spectra_x} illustrates this behavior.
At any given point in time we can find high energy particles in a region closer to the shock front (panel a), while almost no such particles are present in a region further away from the shock (panel b).
However, the reflected thermal particles are present in both regions.
This suggests that reflection occurs already at an early stage in the simulation, giving the reflected particles enough time to reach the region further away from the shock.
The energetic particles, however, could not yet cover the same distance, since they had been released from the shock region at a later time, i.e.~after having been accelerated to high energies.

\ref{app:electrons} contains plots similar to Figs.~\ref{fig:upstream_spectra_t} and~\ref{fig:upstream_spectra_x} which show the energy spectra of the electrons in the upstream plasma.

\section{Summary and Conclusions}
\label{sec:conclusions}


We have investigated the formation of a quasi-parallel shock and the heating and acceleration of particles in an environment resembling the solar wind using fully kinetic PiC simulations.
In three separate simulations we have added He$^{+2}$, He$^{+}$, and C$^{6+}$ ions to our electron-proton plasma.
Although the physical abundances of these heavier ions are low, we have chosen the same number of numerical particles as for protons and electrons to be able to retrieve detailed statistics of all particle species.

In Sect.~\ref{sec:results_sub2} we presented the energy spectra of the ions in the downstream plasma in Fig.~\ref{fig:downstream_spectra}.
The protons show similar behavior in all three simulations, independent of the choice of the additional species of heavy ions.
A thermal spectrum is produced, which then rolls over to a supra-thermal tail at high energies.
However, due to the limited runtime of the simulations, no power law has formed, yet.

The helium and carbon ions also form a thermal spectrum with higher temperatures than the protons.
In fact, the temperatures scale with the ion mass $A$ in units of the proton mass, since the kinetic energy of the incoming upstream ions also scales with their mass.
Similar to the protons, the heavier ions also form a supra-thermal tail at high energies -- although no power law is established.
The maximum energy reached by the particles scales with the charge $Z$ in units of the proton charge, as would be expected from theoretical considerations in the framework of diffusive shock acceleration.

We then study the energy composition and the particle spectra in the upstream in Sect.~\ref{sec:results_sub3}.
In the frame of reference in which the simulations are carried out the downstream is at rest and the upstream flows towards the shock front.
We find that this bulk flow slows down over time (see Figs.~\ref{fig:energy_composition} and~\ref{fig:slowdown}).
The rate of the slowdown changes with the distance to the shock front, suggesting that close to the shock the upstream flow interacts with the particles escaping the downstream.
However, a slowdown occurs also at larger distances from the shock.
The bulk kinetic energy of the upstream is not transferred to thermal or field energy, but we assume that it is instead lost to the reflected and accelerated particles coming from the downstream.

Comparing the energy content of the unperturbed upstream to the energy content of the reflected particles shows that these particles gain about 12-15\% of the upstream energy.
However, the reflected particles do not form a single population.
Instead we find one population that is reflected at the shock front and returns into the upstream while maintaining its thermal energy spectrum.
At higher energies a second population of accelerated particles is formed (see Fig.~\ref{fig:upstream_spectra_t}).
This second population forms a narrow feature in the energy spectrum.

Furthermore Fig.~\ref{fig:upstream_spectra_x} shows that the more energetic population of reflected particles is produced later in the simulation, suggesting that the acceleration takes time.
This is consistent with the idea that the particles cross the shock several times, gaining energy until they can finally escape into the upstream.
The reflected thermal population on the other hand emerges immediately after contact with the shock front.


In this work we have analyzed the dynamics of shock formation and particle acceleration in a fully kinetic model, using PiC simulations.
This approach is clearly different from other plasma models, such as the hybrid model, where the kinetic effects of the electrons are neglected.
Instead of a pair plasma or a pure electron-proton plasma -- which are typically considered when using a PiC code -- we choose to also include heavier ions, such as helium or carbon.
One might argue that PiC is not suitable to study ion dynamics, since the relevant time scales can hardly be resolved due to the enormous computational effort.
For example, the energy spectra of the downstream ions shown in Fig.~\ref{fig:downstream_spectra} do not show the expected power law tail at high energies due to the limited runtime of the simulations.
However, we still think that our work presents an interesting and important numerical experiment.
Our basic results can confirm the findings from hybrid simulations and can thus help to crosscheck different numerical methods.
This is a crucial procedure to improve the reliability of and the trust in numerical experiments in general.

Besides this rather universal result we can now concentrate on different aspects of the dynamics leading to particle acceleration.
Some new results may be drawn from this work already, while others will have to be extracted from future studies concentrating on the details of some individual topics.

As some of our key findings we see that the acceleration of protons and heavier ions proceeds at different time scales and that the ion temperature in the downstream scales with the ion mass $A$.
While this is not surprising it is still interesting to note that this effect can be recovered with a fully kinetic simulation approach and an artificial mass ratio of \mbox{$m_\mathrm{p} / m_\mathrm{e} = 64$}.
In this case, the important parameter is not the ratio of proton to electron mass, but the correct ratio of ion to proton mass.
If the latter is set to the natural value, the correct physical behavior can be recovered.
However, care has to be taken when choosing $m_\mathrm{p} / m_\mathrm{e}$:
With a ratio below \mbox{$m_\mathrm{p} / m_\mathrm{e} = 25$} electrons and ions do not fully decouple.
The shock development is then mostly determined by the electron dynamics instead of the protons.
We will discuss this topic in more detail in an upcoming publication.

Another interesting result is the slowdown of the upstream plasma.
The upstream flow seems to interact with reflected particles in front of the shock.
This interaction region stretches out over several hundred proton inertial lengths.
The velocity profile shown in Fig.~\ref{fig:slowdown}\,b) clearly shows that there is spatial variation in the rate of the slowdown.
To our knowledge this effect has not been studied in detail, yet.
It would be interesting to compare our results with hybrid simulations or to prepare a study which investigates the influence of the shock angle or Mach number on the slowdown.

We have shown that the PiC approach can be used to simulate shock formation and particle acceleration in a plasma containing heavier ions.
With a reasonable choice of the mass ratio $m_\mathrm{p} / m_\mathrm{e}$ the correct physical behavior of all ion species can be retrieved.
However, the computational effort is still high and in the long run a hybrid kinetic approach is, of course, more economic.
PiC simulations on the other hand may be used to further investigate the behavior of the electrons in plasmas plasmas containing different ion species.
Such a study might be a worthwhile project to investigate the mechanism leading to the different abundances of high energy electrons seen in Fig.~\ref{fig:downstream_spectra_elec}.

\section*{Acknowledgments}
We acknowledge the use of the \emph{ACRONYM} code and would like to thank the developers (Verein zur F\"orderung kinetischer Plasmasimulationen e.V.) for their support.

We acknowledge support by the Max Planck Computing \& Data Facility (MPCDF), which provided the computational resources used for the simulations presented in this work.

CS would like to thank J\"org B\"uchner for having him as a guest at the MPS and the Max-Planck-Society for granting a stipend.

PK and FS acknowledge support from the NRF and DST of South Africa by the following disclosure:
This work is based upon research supported by the National Research Foundation and Department of Science and Technology.
Any opinion, findings and conclusions or recommendations expressed in this material are those of the authors and therefore the NRF and DST do not accept any liability in regard thereto.

FS acknowledges support from the Deutsche Forschungsgemeinschaft through grant SP1124/9.

PM acknowledges funding by the Max-Planck-Princeton Center for Plasma Physics.

\appendix
\setcounter{figure}{0} 
\section{Electron Spectra}
\label{app:electrons}

For completeness we show the energy spectra of the electrons in simulations S1, S2, and S3 in this appendix.

\begin{figure}[htb]
      \centering
      \includegraphics[width=0.8\linewidth]{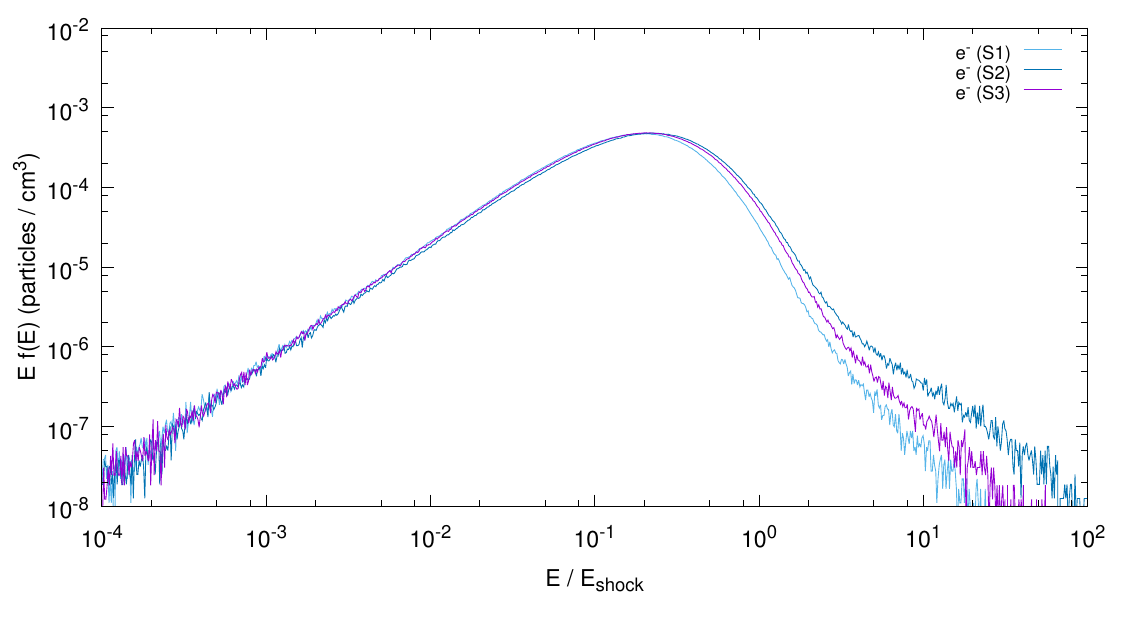}
      \caption{
      Downstream energy spectra of electrons from simulations S1, S2, and S3 at $t \, \Omega_\mathrm{p,u}= 68.0$.
      The energy is normalized to the shock energy $E_\mathrm{shock} = 0.5 \, m_\mathrm{p} \, v_\mathrm{u}^2$.
      }
      \label{fig:downstream_spectra_elec}
\end{figure}

Figure~\ref{fig:downstream_spectra_elec} shows the energy spectra of the downstream electrons measured in the rest frame of the downstream, similar to Fig.~\ref{fig:downstream_spectra}.
We find a thermal population which includes most of the electrons and which is mostly the same for all three simulations.
At high energies ($E/E_\mathrm{shock} > 2$) a supra-thermal population emerges.
The number of particles contributing to this high-energy population varies with the simulation setup.
It is interesting to note that the number of supra-thermal particles seems to be correlated with the electron temperature (i.e. the position of the peak of the spectrum):
Simulation S2 produces the most energetic particles and the highest temperature of the downstream electrons.
The temperature is slightly lower in S3 and still lower in S1.

At first thought one might assume that the different heavier ions (He and C) are responsible for the differences in the high energy tails of the electron spectra.
However, this is unlikely, since He and C produce field and density fluctuations on very large scales (see Fig.~\ref{fig:phase_space}), while the electrons react to fluctuations at much smaller scales.
The results of \cite{romanky_2017} support this claim:
They investigate the effect of an additional population of helium ions on the shock acceleration characteristics of protons and electrons.
While protons are accelerated more efficiently, the electron spectra remain unaffected by the heavier ions.
This is even more interesting, considering the low mass ratio of $m_\mathrm{p} / m_\mathrm{e} = 20$ they chose.

The shape of the high-energy tails of the spectra in Fig.~\ref{fig:downstream_spectra_elec} suggest that the cutoff energy lies above $E / E_\mathrm{shock} = 100$.
However, we are not able to follow the spectrum to higher energies.
Due to the limited number of numerical particles in the simulation the spectrum reaches a hard cutoff slightly above the energy range shown in the plot:
There are simply no numerical particles at such high energies.
For future investigations of electron acceleration a greater number of numerical particles per cell would have to be chosen.

Next we present the energy spectra of the electrons in the upstream.
Data is gathered as described in Sect.~\ref{sec:results_sub3} for the protons and heavier ions.

\begin{figure}[h!]
      \centering
      \includegraphics[width=0.5\linewidth]{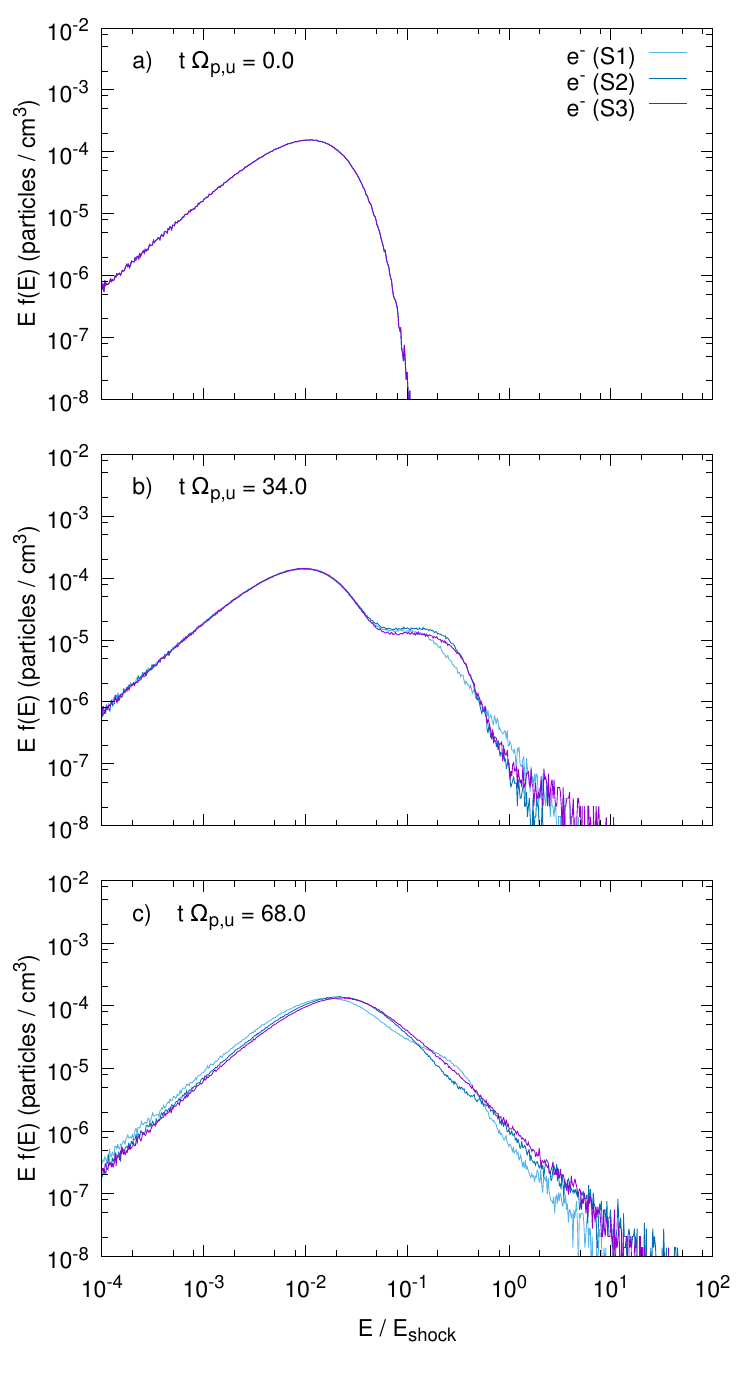}
      \caption{
      Upstream energy spectra of the electrons from simulations S1, S2, and S3 at various points in time as labeled in the panels.
      Particle data was gathered in a spatial interval between $159.0 \, \lambda_\mathrm{p,u}$ and $238.5 \, \lambda_\mathrm{p,u}$ from the shock front.
      The energy is given in the rest frame of the unperturbed upstream.
      }
      \label{fig:upstream_spectra_t_elec}
\end{figure}

Figure~\ref{fig:upstream_spectra_t_elec} shows the spectra of electrons in the upstream at three different points in time but at the same distance from the shock front, similar to Fig.~\ref{fig:upstream_spectra_t}.
As for the ions, the electrons form a thermal spectrum in the unperturbed upstream (e.g. at $t=0$, see panel a).
As the shock forms and particles are reflected and accelerated, a second component is added to the energy spectra.
Reflected electrons form a second peak at higher energies than the thermal peak (see panel b).
Unlike for the protons and the heavier ions this second peak is much closer to the first one.
With an increasing number of reflected and accelerated electrons escaping from the shock at later times, the spectra form a relatively smooth slope towards high energies (see panel c).
The second peak is then no longer visible.
Interestingly the narrow feature at highest energies that was observed for the ions in Fig.~\ref{fig:upstream_spectra_t_elec} is not reproduced for the electrons.

\begin{figure}[h!]
      \centering
      \includegraphics[width=0.5\linewidth]{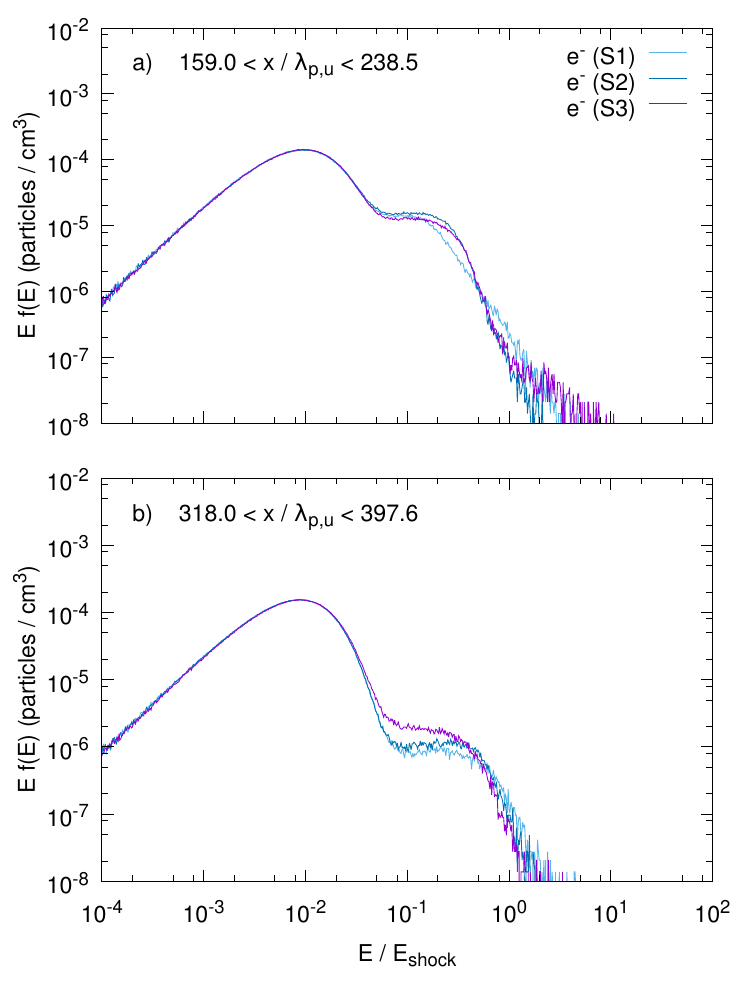}
      \caption{
      Upstream energy spectra of the electrons from simulations S1, S2, and S3 at $t \, \Omega_\mathrm{p} = 34.0$ and two different distances from the shock front as labeled in the panels.
      }
      \label{fig:upstream_spectra_x_elec}
\end{figure}

The maximum energy reached by the electrons increases over time (as is the case for the ions).
The acceleration process takes place close to the shock front, leading to spatial variations in the spectrum as the particles escape further into the upstream.
We show this spatial dependence in Fig.~\ref{fig:upstream_spectra_x_elec}, which shows electron spectra at two different distances from the shock but at the same point in time (similar to Fig.~\ref{fig:upstream_spectra_x}).

\section*{References}
\bibliography{ref}

\end{document}